\documentclass[seceq]{ptptex}
\usepackage{graphicx}
\newcommand{\be}{\begin{equation}}
\newcommand{\ee}{\end{equation}}

\newcommand{\bea}{\begin{eqnarray}}
\newcommand{\eea}{\end{eqnarray}}
\newcommand{\dis}{\displaystyle}

\title{%        %You can use \\ for explicit line-break.
Analysis of Realized Volatility in Two Trading Sessions of the Japanese Stock Market
}

\author{%       %Use \scshape for the family name.
Tetsuya \textsc{Takaishi}$^{1,}$\footnote{E-mail:tt-taka@hue.ac.jp}%
,
Ting Ting \textsc{Chen}$^{2}$
and 
Zeyu \textsc{Zheng}$^{3}$
}
\inst{%     %Affiliation, neglected when [addenda] or [errata].
$^{1}$Hiroshima University of Economics, Hiroshima 731-0192, Japan\\
$^{2}$Faculty of Integrated Arts and Sciences, Hiroshima University, Higashi-Hiroshima 739-8521, Japan\\
$^{3}$Department of Environmental Sciences, Tokyo University of Information Sciences, Chiba 265-8501, Japan
}

\abst{%         %This abstract is neglected when [addenda] or [errata].
We analyze realized volatilities constructed
using high-frequency stock data on the Tokyo Stock Exchange.
In order to avoid non-trading hours issue in volatility calculations
we define two realized volatilities
calculated separately in the two trading sessions of the Tokyo Stock Exchange,
i.e. morning and afternoon sessions.
After calculating the realized volatilities at various
sampling frequencies we evaluate the bias from the microstructure noise
as a function of sampling frequency.
Taking into account of the bias to realized volatility
we examine returns standardized by realized volatilities and
confirm that price returns on the Tokyo Stock Exchange
are described approximately by Gaussian time series with
time-varying volatility, i.e. consistent with a mixture of distributions hypothesis.
}

\begin{document}

\maketitle

\section{Introduction}

In recent years statistical properties of asset price returns 
have been extensively studied in econophysics\cite{B1,B2,B3,B4,CONT}.
One of the pronounced properties is that
the probability distribution of returns exhibits
a fat-tailed distribution which is not Gaussian\cite{Mandelbrot,Fama,Mantegna,Gabaix}.
It is revealed that the tail distributions of returns
exhibit the power law\cite{Stanley1,Stanley2,Stanley3} and
the distributions of returns at short time scale are well fitted by
Student's t-distributions\cite{Praetz,ST1,ST2,ST3,GERIG} which is also known as 
the q-Gaussian distributions in nonextensive statistical mechanics\cite{Tsallis,Tsallis2}. 
It is crucial to understand the price dynamics which results in the fat-tailed distributions.
A possible explanation for the origin of the fat-tailed distributions
is that the  price dynamics follows 
the Gaussian process with time-varying volatility,
which is called the mixture of
distributions hypothesis(MDH)\cite{Clark}.
With this hypothesis  each return $r_t$ at time $t$
is described by $r_t=\sigma_t \epsilon_t$, where $\sigma_t^2$ is
a variance of the Gaussian distribution and $\epsilon_t$ is a Gaussian random variable
with mean 0 and variance 1.
From the MDH the return distributions are derived as 
the superposition of the conditional return distribution with $\sigma_t$ and
the distribution of volatility $\sigma_t^2$.  
This view is closely related to the superstatistics\cite{Beck}
where the statistics of physical systems are separated by two time scales and 
unconditional distributions are obtained by the superposition of these statistics.

In the MDH the shape of return distributions is determined by 
the volatility distributions which are not known a priori.
Empirical studies suggested that the volatility distributions
are described by the inverse gamma distribution or log-normal distribution\cite{Praetz,RV2,RV3,Beck2,GERIG,Takaishi}
with which the unconditional return distributions result in fat-tailed distributions.
Especially the inverse gamma distribution gives the Student's t-distribution (q-Gaussian distribution) 
for the unconditional return distribution.

The MDH itself does not derive the shape of volatility distributions. 
The volatility dynamics which accounts for the shape of volatility distributions 
may depend on microscopic features of the markets  
such as volume, transactions, information arrival etc.\cite{V1,V2,V3,V4} which we do not address here. 
Toward the complete dynamics of price, 
it is important to check whether the MDH holds for the real market price data. 

The consistency between the MDH and the real markets data
can be checked by the standard normality of the returns standardized by volatility.  
Namely under the MDH the standardized returns given by $r_t/\sigma_t$ should show 
Gaussian-distributed variables with variance ( or standard deviation ) 1 and kurtosis 3.
Note that  it is crucial to check both properties of variance 1 and kurtosis 3.
In order to perform this check we need to estimate the value of volatility
since volatility is not a direct observable from the market data. 
The conventional and popular approach for volatility estimation
is to use parametric models, such as GARCH-type\cite{ARCH,GARCH} and stochastic volatility models\cite{SVMCMC1,SV,SVMCMC2} which 
are designed to capture the relevant properties observed in financial time series, e.g.
volatility clustering and fat-tailed distribution for returns. 
Each model is constructed under  specific assumptions for the time-varying volatility process.
The volatility measures obtained from those models could differ each other.
Thus the validity for volatility measures is also another issue which has to be considered.

In this paper in order to deal with more accurate volatility we focus on realized volatility constructed 
as a sum of intraday squared returns\cite{RV1,RV2,RV3} and examine the view of the MDH for 
stock returns on the Tokyo Stock Exchange.
The recent availability of high-frequency financial data enables us to 
easily access to calculations of realized volatility and an advantage over parametric models 
is that realized volatility is a model-free estimate.
Furthermore if there is no measurement error this estimate 
provides an unbiased volatility measure for the integrated volatility
and converges to the integrated volatility in the limit of infinite high sampling frequency.
However actual market prices suffer from measurement errors caused by 
microstructure noise such as bid-ask spread etc. \cite{Campbell}.
Therefore in practical calculation realized volatility is considered to be biased by such unwanted microstructure noise.
Provided that the log-price observed in the market is contaminated with an independent noise\cite{Zhou}
we expect that this bias will be especially crucial at high frequency and 
increases with increasing the sampling frequency.
Such behavior has been actually observed and 
can be depicted in the so called "volatility signature plots"\cite{VSP}.
In order to reduce this bias and also to maintain accuracy of the measurement
typically the returns used for realized volatility calculations 
are sampled at a moderate frequency, e.g. 5-min. frequency\cite{RV2,RV3,Blair,Koopman,Ghysels}.

When we deal with the daily volatility estimated as realized volatility 
another problem arises, that is 
non-trading hours issue in measuring daily volatility.
In the Tokyo Stock Exchange, 
there are two trading sessions in a day,
i.e. Morning Session (MS) 9:00-11:00 and 
Afternoon Session (AS) 12:30-15:00.
Thus available stock price data are taken only in the two sessions and
outside of these sessions no available data exist.
If the realized volatility is calculated with data only from trading hours, 
the resulting measure could be biased and unreliable as daily volatility.
To circumvent this problem 
Hansen \& Lunde\cite{Hansen} proposed to introduce an adjustment factor
which corrects the realized volatility so that  the average of the realized volatility matches the variance of  daily returns.

The standard normality of standardized return time series has been examined by using realized volatility\cite{Zhou,RV2,RV3,RV4,Takaishi} 
and it is found that the distributions of the standardized return series are approximately described by 
Gaussian distributions.
However  from the analysis of stocks in the Dow Jones Industrial Average (DJIA) the standard deviations of the standardized return series  turned out to be 
still apart from 1 and all the values are less than 1. 
Moreover the values of the kurtosis also vary significantly around 3. 

Instead of using log-price returns Ref.\cite{AN} examines  the price change for U.S. stocks and finds that 
while  the standardized price changes are well described by Gaussian-distributed random variables
the kurtosis exceeds 3. In addition the variance of the the standardized price changes has not been 
examined. Thus it has not been concluded that the MDH exactly holds for the price changes.

From previous studies it is turned out that the return dynamics is approximately 
consistent with the MDH. 
However the origin of the deviations from the MDH which appear in variance and kurtosis are 
not known yet. 
In order to reveal the origin and examine whether the MDH precisely holds for the returns 
it is important to control the bias in measuring volatility. 
In this paper we carefully evaluate the bias from the microstructure noise remaining 
in the realized volatility  
and examine the standard normality of standardized returns and the consistency with the MDH.

Under the MDH the standardized returns should exhibit properties of variance 1 and kurtosis 3, i.e. standard normality.
In order to examine both properties on the Tokyo Stock Exchange
we have to avoid "non-trading hours" issue, i.e. introduction of the Hansen \& Lunde (HL) adjustment factor 
to the realized volatility. 
Otherwise the realized volatility with the HL adjustment factor can change the values of variance and  
we can not check  the property of variance 1 under the MDH.
In Ref.\cite{Takaishi} standardized daily returns on the Tokyo Stock Exchange have been analyzed 
and it is found that  the standardized daily returns exhibit the Gaussianity, i.e. kurtosis 3. 
However due to the "non-trading hours" issue  the property of variance 1 which is needed for 
the consistency check of the MDH has not
been confirmed yet.
To avoid "non-trading hours" issue here we do not deal with
a whole day volatility. Instead  we calculate
two realized volatilities separately  in the two trading periods, i.e.
one measured in the MS and the other in the AS.
In this way we obtain two realized volatilities $RV_{MS}$ and $RV_{AS}$ without the HL adjustment factor.
By investigating standardized returns in the MS and AS separately
we are able to examine both properties of variance 1 and kurtosis 3, i.e. the MDH.

The outline of the paper is as follows. Section 2 describes the realized volatility, 
the microstructure noise in the realized volatility and the HL adjustment factor.
In Section 3 we explain the data on the Tokyo Stock Exchange analyzed in this study.
In Section 4 we analyze the microstructure noise effects on the realized volatility.
In Section 5 the MDH is examined by the returns standardized by realized volatility. 
We conclude in Section 6 with a brief summary.

\section{Realized volatility}

The realized volatility is a model-free estimate of volatility 
constructed as a sum of squared returns\cite{RV1,RV2,RV3}.
Let us assume that the logarithmic price process $\ln p(s)$ follows 
a continuous time stochastic diffusion,
\be
d\ln p(s) =\tilde{\sigma}(s)dW(s),
\label{eq:SD}
\ee
where $W(s)$ stands for a standard Brownian motion  and $\tilde{\sigma}(s)$ is a spot volatility at time $s$.  
In financial applications our main interest is to measure an integrated volatility which is defined by  
\be
\sigma_h^2(t) =\int_{t}^{t+h}\tilde{\sigma}(s)^2ds,
\label{eq:int}
\ee
where $h$ stands for the interval to be integrated.
If we consider daily volatility $h$ takes one day.
Since $\tilde{\sigma}(s)$ is latent and not available from market data,
eq.(\ref{eq:int}) can not be evaluated analytically.

Constructing $n$ intraday returns  from high-frequency data,
the realized volatility $RV_t$ is given by a sum of squared intraday returns, 

\be
RV_t =\sum_{i=1}^{n} r_{t+i\Delta}^2,
\label{eq:RV}
\ee
where $\Delta$ is a sampling period\footnote{Small sampling period corresponds to high sampling frequency.}
defined by $\Delta=h/n$ and
returns are given by log-price difference,
\be
r_{t+i\Delta}=\ln P_{t+i\Delta}-\ln P_{t+(i-1)\Delta}.
\ee

Without any bias 
$RV_t$ goes to the integrated volatility of eq.(\ref{eq:int}) in the limit of $n \rightarrow \infty$. 
In the presence of bias such as microstructure noise\cite{Campbell},
the convergence of $RV_t$ to the integrated volatility is not guaranteed.
Following Zhou\cite{Zhou} we assume that the log-price observed in financial markets is contaminated with independent noise, i.e.
\be
\ln P^{*}_t =  \ln P_t +\xi_t,
\ee
where $\ln P_t^{*}$ is the observed log-price in the markets which consists of the true log-price $\ln P_t$ and
noise $\xi_t$ with mean 0 and variance $\omega^2$. 

Using this assumption the observed 
return $r^{*}_t$ is given by 
\be
r^{*}_t=r_t +\eta_t,
\ee
where $\eta_t=\xi_{t}-\xi_{t -\Delta}$.
Thus $RV^{*}_t$ which is actually observed from the market data is obtained 
as a sum of the squared returns $r^{*}_t$, 
\bea
RV_t^{*}& = &\sum_{i=1}^{n} (r^{*}_{t+i\Delta})^2,  \\
   & =& RV_t + 2\sum_{i=1}^n r_{t+i\Delta}\eta_{t+i\Delta} + \sum_{i=1}^n \eta_{t+i\Delta}^2.
\label{eq:rvnoise}
\eea
With these independent noises  the bias appears as $\sum_{i=1}^n \eta_{t+i\Delta}^2$
which corresponds to $\sim 2n\omega^2$.
Thus due to the bias the $RV^{*}_t$ diverges as $n \rightarrow \infty$.
Such divergent behavior has been seen in the volatility signature plot\cite{VSP}.
for liquid assets.

To avoid distortion from microstructure noise a good sampling frequency which
reduces the bias but maintains accuracy of the realized volatility measure should be considered.
It is suggested that 5-min sampling frequency is short enough for realized volatility calculation\cite{RV2,RV3}.
In this paper we also use 5-min sampling frequency to analyze standardized returns.

When we consider daily realized volatility 
we have to cope with another problem which is that 
high-frequency data are not available for the entire 24 hours
except for some exchange rates.
At the Tokyo stock exchange market domestic stocks are traded in
the two trading sessions: (1)morning trading session (MS) 9:00-11:00.
(2)afternoon trading session (AS) 12:30-15:00.
The daily realized volatility calculated without including intraday returns during
the non-traded periods
can be underestimated.

Hansen and Lunde\cite{Hansen} advocated an idea to circumvent the problem by
introducing an adjustment factor which 
modifies the realized volatility so that the average of the realized volatility matches
the variance of the daily returns.
Let $(R_1,...,R_N)$ be $N$ daily returns.
The adjustment factor $c$ is given by
\be
c=\frac{\sum_{t=1}^{N}(R_t-\bar{R})^2}{\sum_{t=1}^{N}RV_t},
\ee
where $\bar{R}$ denotes the average of $R_t$.
Then using this factor the daily realized volatility is modified to $cRV_t$.
When we analyze the returns standardized by the realized volatility
this adjustment factor largely affects the standard deviation of 
the standardized returns.

In order to avoid the non-trading hours issue
and the introduction of the adjustment factor to the realized volatility
we consider two realized volatilities:(i) $RV_{MS}$, realized volatility in the morning session and
(ii) $RV_{AS}$, realized volatility in the afternoon session.
Since these realized volatilities are calculated separately 
and each does not include non-trading hours, 
no adjustment factor is needed. 

\section{Data Analyzed}

Our analysis is based on data for 5 stocks, 1:Mitsubishi Co., 2:Nomura Holdings Inc., 3:Nippon Steel, 4:Panasonic Co. and 5:Sony Co.
These stocks are listed in the Topix core 30 index which 
includes liquid stocks of the Tokyo Stock Exchange.
Our data set begins June 3, 2006 and ends December 30, 2009. 

\begin{figure}[ht]
\vspace{12mm}
\centering
\includegraphics[height=8.5cm]{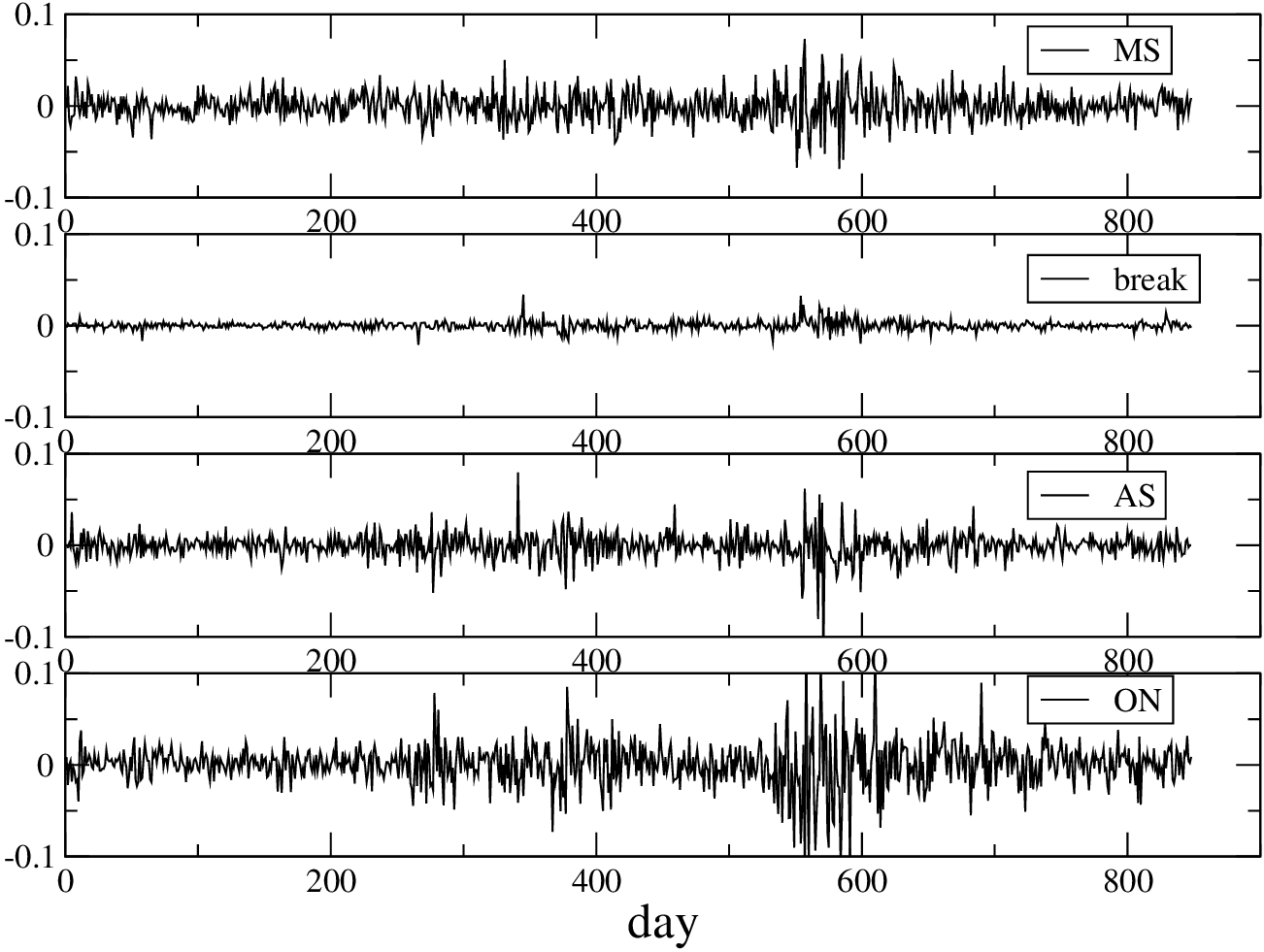}
\caption{
Return time series in the different time zones for Mitsubishi Co. as representative series.}
%\vspace{1mm}
\label{fig:Rt-mitsubishi}
\end{figure}

Fig.1 shows  daily intraday return time series in the different time zones of the Tokyo Stock Exchange.
The figure shows the time series of Mitsubishi Co. as a representative one.  
Each return is calculated by the log-price difference 
between the opening and closing prices of the corresponding zone.
For instance $R_{MS,t}$ is given by $\ln P_{MS,t}^o - \ln P_{MS,t}^c$, where 
$P_{MS,t}^o$ is the opening price of the morning session on the day $t$ and 
$P_{MS,t}^c$ the closing price of the morning session on the day $t$.
In a similar manner $R_{AS,t}$ is given by $\ln P_{AS,t}^o - \ln P_{AS,t}^c$.
Returns in the lunch break are calculated by
$R_{break,t}=\ln P_{MS,t}^c - \ln P_{AS,t}^o$.
$R_{ON,t}$ is the overnight return given by $\ln P_{MS,t}^o - \ln P_{AS,t-1}^c$.
We see that the magnitude of the returns in the lunch break is very small which means 
that the price change in this zone is small.
On the other hand in the overnight zone the magnitude of the returns 
is big as well as in the trading zones.

We focus on two realized volatilities  which are constructed using data from different trading zones.
Let us denote $RV_{MS,t}$ (  $RV_{AS,t}$ ) as the realized volatility calculated using data in the MS (AS).
For instance  $RV_{MS,t}$ is defined by
\be
 RV_{MS,t}=\sum_{i=1}^n r_{t_{MS}+i\Delta_{MS}}^2,
\label{eq:RVMS}
\ee
where $n$ and $\Delta_{MS}$  stand for the sampling number and sampling period respectively, and
the relation between $n$ and $\Delta_{MS}$ is given by $\Delta_{MS}=h_{MS}/n$, 
where $h_{MS}$ is the trading time of the MS, i.e. 120 min. at the Tokyo stock exchange market.
$t_{MS}$ is the opening time of the MS, i.e. 9:00.
$RV_{AS,t}$ is also defined in a similar manner, e.g. $h_{AS}=$ 150 min. and $t_{AS}=$ 12:30.

\section{Microstructure noise}
In order to quantify the microstructure noise
we measure the realized volatility  at various sampling frequencies and
make "volatility signature plot"\cite{VSP} to visualize the bias effect from
the microstructure noise.
Fig.2 shows the volatility signature plot for Mitsubishi Co. as a representative one.
The top (bottom ) panel of the figure shows the average realized volatility in the MS (AS)
as a function of sampling period $\Delta$.  
As seen in Fig.2  due to the microstructure noise 
the average realized volatility increases with increasing 
of the sampling frequency ( with decreasing of the sampling period $\Delta$).
We also find quantitatively similar results for volatility signature plots of other stocks.
The average realized volatility is well fitted to a functional form of $\dis a_0(1+a_1/\Delta)$ 
which is an expected from eq.(\ref{eq:rvnoise}).
The dotted lines in Fig.2 show the fitting results.
The true average realized volatility can be evaluated to be $a_0$.

The value of $\dis \delta(\Delta)=a_1/\Delta$ corresponds to the bias from the microstructure noise at
the sampling period $\Delta$.
The fitted parameters $a_0,a_1$ and the bias values at $\Delta=5min$ are summarized in Table I.
It is found that the bias in the afternoon session is larger than the one in the morning session.
It seems that the average realized volatility stabilizes around 5-10 min. within error bar.
However if we compare the bias to the true average volatility which is evaluated as $a_1/\Delta$
the bias at 5 min. still contributes largely.
Especially in the afternoon session, the bias at 5 min. proceeds 30\% for most stocks examined here. 
These biases result in decreasing the variance of standardized returns.
Later we incorporate these biases to calculate the variance of standardized returns. 

\begin{figure}[ht]
\vspace{13mm}
\centering
\includegraphics[height=8.0cm]{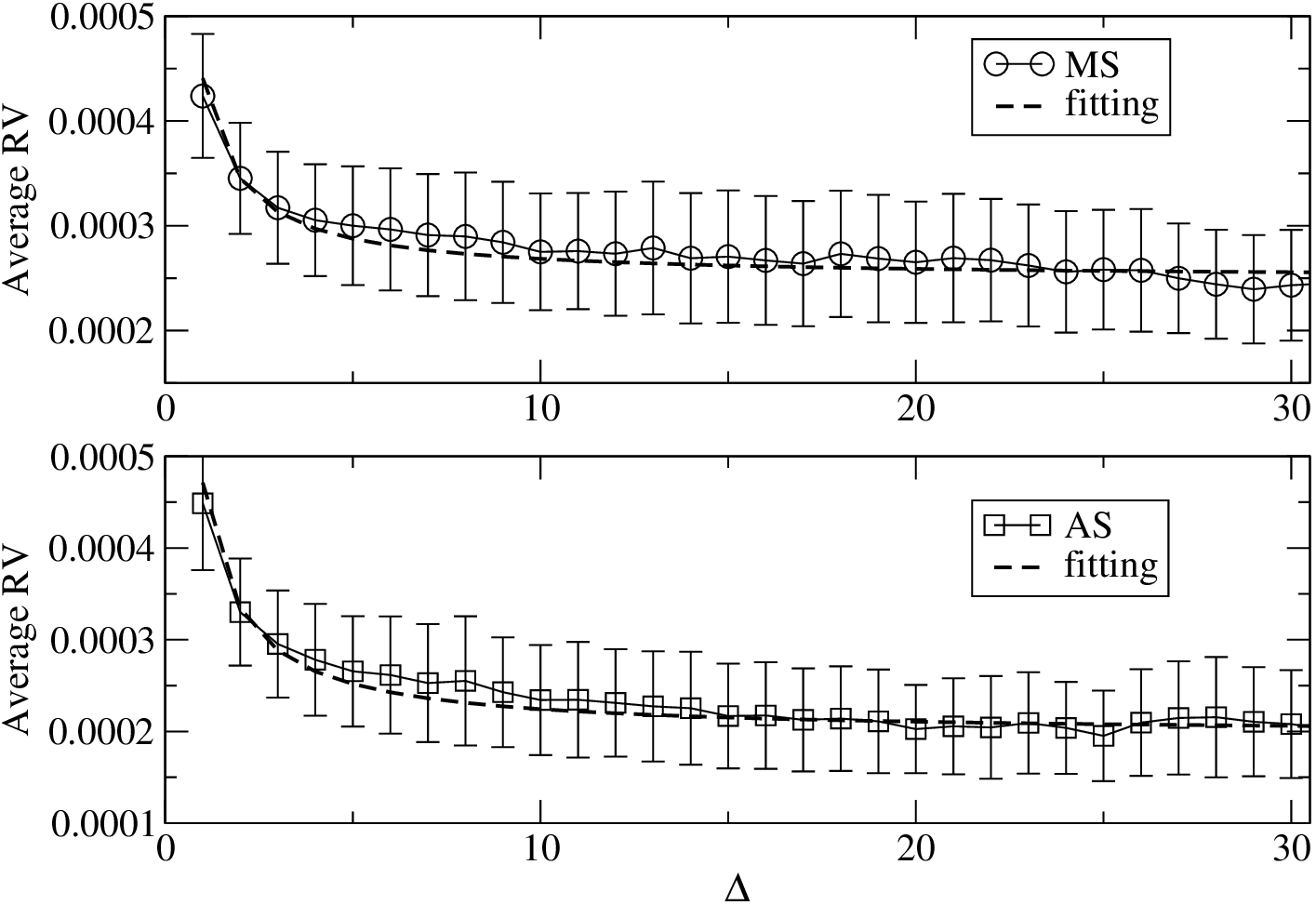}
\caption{
Average realized volatility in the MS (top) and AS (bottom) for Mitsubishi Co. 
as a function of sampling period $\Delta$ (min). The dotted lines show the fitting results 
to the expected formula of $a_0(1+a_1/\Delta)$.}
%\vspace{1mm}
\label{fig:rvave-mitsubishi}
\end{figure}

\begin{table}[t]
  \centering
  \caption{Fitted parameters and bias at 5min. $a_0$ corresponds to the average RV in the limit of $n \rightarrow \infty$.
The bias at $\Delta=$5min. is calculated by $\delta= a_1/\Delta$.}
  \label{tab:1}
    \begin{tabular}{cc|ccccc}

\hline
Stock      &          & 1:Mitsubishi       &  2:Nomura          & 3:Nippon St.           & 4:Panasonic        & 5:SONY              \\
\hline
MS & $a_0$           & $2.5\times10^{-4}$ & $2.5\times10^{-4}$ & $2.0\times10^{-4}$ & $1.6\times10^{-4}$ & $1.7\times10^{-4}$  \\
   & $a_1$           & 0.705              & 0.366              & 1.09               & 0.643              & 0.780               \\
   & $\delta(5min.)$ & 0.141              & 0.073              & 0.228              & 0.126              & 0.156               \\
AS & $a_0$           & $1.8\times10^{-4}$ & $1.8\times10^{-4}$ & $1.7\times10^{-4}$ & $1.1\times10^{-4}$ & $1.2\times10^{-4}$  \\
   & $a_1$           & 1.58               & 0.897              & 1.88               & 1.59               & 1.80                \\
   & $\delta(5min.)$ & 0.316              & 0.179              & 0.376              & 0.318              & 0.360               \\
\hline 
    \end{tabular}
\end{table}

\section{Standard normality of standardized returns}
From the MDH 
return time series, $R_{MS,t}$ and $R_{AS,t}$ are expected to be 
\be
R_{MS,t}=\sigma_{MS,t}\epsilon_t,
\label{eq:Rt1}
\ee
and
\be
R_{AS,t}=\sigma_{AS,t}\epsilon_t,
\label{eq:Rt2}
\ee
respectively.
$\sigma_{MS,t}^2$ ($\sigma_{AS,t}^2$) is an integrated volatility in the morning ( afternoon ) session.
$\epsilon_t$ is an independent Gaussian random variable with mean 0 and variance 1.  
Substituting realized volatility for the integrated volatility, i.e. $\sigma_{MS,t}=RV_{MS,t}^{1/2}$ and $\sigma_{AS,t}=RV_{AS,t}^{1/2}$ 
we expect that returns standardized by those standard deviation
exhibit Gaussian-distributed time series.   

Table II (III) shows the standard deviations and kurtoses of original and standardized returns in the MS (AS).
We find that the kurtoses for the original returns are very high, compared to 3 of Gaussian random variables.  
On the other hand the kurtoses of the standardized returns come near 3.
However it is observed that in the MS all kurtoses are slightly smaller than three. 
At the present moment we do not know what causes this slight difference between the MS and AS. 
We also tested the normality by the Anderson-Darling test and 
the normality was not rejected ( p-values are listed in Tables ). 

We find that the standard deviation of the standardized returns also approaches one. 
However in most cases the values of the standard deviation are slightly less than one.
This difference could be explained by 
the bias which still remains at 5min. sampling frequency.
In Tables II and III we also list the bias-corrected standard deviation.
Let $\sigma$ be the standard deviation of the standardized returns.
The bias-corrected standard deviation is given by $\sigma*(1+\delta(5min))^{1/2}$,
where $\delta(5min)$ is the bias at 5-min sampling frequency listed in Table I. 
It is evident that the bias-corrected standard deviation comes more close to one. 
Although the values of the bias-corrected  standard deviation in the AS  for some stocks 
appear to be slightly less than 1,  
other results turns out to be satisfactory very close to one, 
compared to the average value of 0.8 which is obtained from the standardized returns of the DJIA stocks\cite{RV3}.

\begin{table}[t]
  \centering
  \caption{Standard deviation and kurtosis of original and standardized returns in the morning session. 
The values of parentheses show statistical errors estimated by the Jackknife method.
AD stands for the Anderson-Darling normality test.
 }
  \label{tab:2}
    \begin{tabular}{cc|ccccc}
\hline
Stock &                             & 1:Mitsubishi & 2:Nomura &3:Nippon St.  &4:Panasonic & 5:Sony    \\
\hline
$R_t$ & std.dv. ($\times 10^{2}$)   & 1.58(20)         & 1.58(16)     & 1.53(17)        &1.28(13)        &1.36(14)        \\
      & kurt.                       & 4.99(50)         & 4.90(71)     & 6.10(100)        &5.53(94)        &4.71(39)        \\
\hline
$R_t/RV_t^{1/2}$ & std.dv. $(=\sigma)$& 0.915(32)        & 0.951(20)    & 0.909(30)       &0.895(26)       &0.921(26)       \\
                 & kurt.            & 2.75(13)         & 2.68(11)     & 2.79(13)        &2.66(12)        &2.66(12)        \\
\hline
    &   $\sigma*(1+\delta(5min))^{1/2}$       & 0.977(35)         & 0.985(31)     & 1.007(33)        &0.950(28)       &0.990(28)        \\
\hline
AD  & p-value & 0.310 & 0.208  & 0.115 & 0.065 & 0.085  \\
\hline
    \end{tabular}
\end{table}

\begin{table}[t]
  \centering
  \caption{Same as in Table II but in the afternoon session.}
  \label{tab:3}

\begin{tabular}{cc|ccccc}
\hline          
Stock                &              & 1:Mitsubishi & 2:Nomura &3:Nippon St. & 4:Panasonic & 5:Sony   \\
\hline
$R_t$ &std.dv. ($\times 10^{2}$)    & 1.40(22)         & 1.34(15)     & 1.50(24)    &1.12(16)         &1.25(21)      \\
      &kurt.                        & 17.3(85)         & 8.66(25)     & 24.8(144)    &11.9(36)        &18.3(61)      \\
\hline
$R_t/RV_t^{1/2}$&  std.dv. $(=\sigma)$ & 0.787(27)     & 0.825(47)    & 0.837(36)   &0.795(33)        &0.820(41)      \\
      &kurt.                        & 3.18(20)         & 3.21(15)     & 3.20(20)    &2.91(14)         &3.16(18)        \\
\hline
&  $\sigma*(1+\delta(5min))^{1/2}$     & 0.903(31)        & 0.896(52)    & 0.982(42)   &0.913(38)        &0.957(48)      \\
\hline
AD  & p-value & 0.159 & 0.063 & 0.087 & 0.266 & 0.196 \\
\hline
    \end{tabular}
\end{table}

As  stylized facts of asset returns it is well known that
while linear autocorrelations of returns  
are not significant except for very small time scale  the autocorrelations of
absolute returns decay very slowly.
On the other hand the standardized returns are expected to be independent Gaussian variables 
and thus not only the  standardized returns but also absolute standardized returns should 
show insignificant autocorrelation. 
Fig.3 compares the autocorrelation function (ACF) between absolute returns and absolute standardized returns\footnote{We have also verified that the standardized returns show no significant autocorrelation.}. 
The top (bottom) panel  shows the ACF in the MS (AS).
The solid lines in the figure show 95\% confidence limits. 
We see that while the autocorrelation function of absolute returns decays very slowly 
the autocorrelation function of absolute standardized returns immediately disappears in the noise level of 95\% confidence band.
These features also support the view of the MDH.

\begin{figure}[ht]
\vspace{15mm}
\centering
\includegraphics[height=8.5cm]{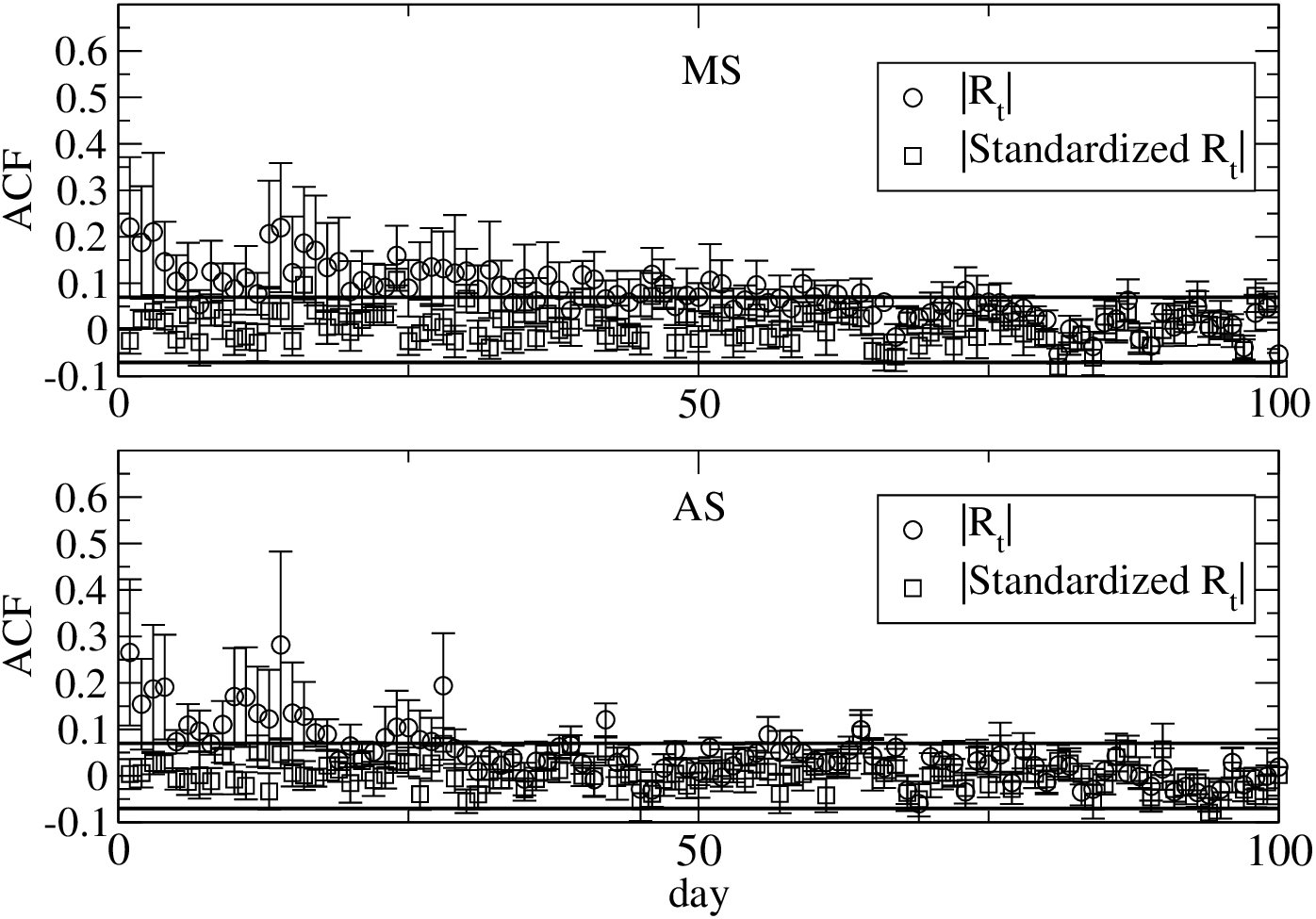}
\caption{
Autocorrelation functions (ACF) of absolute returns and absolute standardized returns
for Mitsubishi Co. The top (bottom)  panel shows the ACF in the MS (AS). 
Solid lines in the figure show 95\% confidence limits.}
%\vspace{1mm}
\label{fig:corrabs-mitsubishi}
\end{figure}

\section{Conclusion}

We constructed two realized volatilities $RV_{MS,t}$ and $RV_{AS,t}$ 
defined in the two trading sessions of the Tokyo Stock Exchange. 
Using the two realized volatilities
we investigated standardized returns in the MS and AS separately and
examine properties of standard deviation 1 and kurtosis 3 under the MDH.  
By calculating the realized volatilities at various sampling frequencies 
we quantified the bias from the microstructure noise as a function of 
the sampling frequency.
Taking into account of the bias effect we find 
that the bias-corrected standard deviations of standardized returns in the MS and AS come close to one. 
Furthermore we also find that the values of the kurtosis in the MS and AS come close to three.
In Ref.\cite{Takaishi} standardized daily returns on the Tokyo Stock Exchange 
have been examined and the normality of the standardized daily returns 
is confirmed. However due to non-trading hours issue 
it was not confirmed that the standard deviation of the standardized daily returns
comes to one.
Our observations confirm that standardized returns in the MS and AS on the Tokyo Stock Exchange 
show both properties of standard deviation 1 and kurtosis 3, expected from the MDH.
Thus we conclude that the price dynamics on the Tokyo Stock Exchange is consistent with the MDH.

While the kurtoses in the AS are consistent with three,  
we observe slightly smaller kurtoses in the MS.
Although  we do not understand this slight difference of the kurtoses between
the MS and AS this might be caused by the
trading duration time difference or trading activity difference 
between the MS and AS. 
We also observe slight deviation from one for standard deviations of standardized returns of some stocks.
The slight deviation might indicate that there still remain other small biases not considered here. 
For instance the diffusion assumption of eq.(\ref{eq:SD})
might need an additional term such as 
the jump component\cite{RV4}. When the jump effect is present it might distort 
the standard normal property. 
In order to fully understand  the price dynamics, 
in future it should be clarified whether such slight differences in the standard deviations and kurtoses
actually indicate any important effects or not.

Under the MDH it is expected that not only the standardized returns but also
the absolute standardized returns show no significant autocorrelation.
We verified that 
the absolute  standardized returns have no autocorrelations. 
These results also support the MDH for the price dynamics of the stocks on the Tokyo Stock Exchange.

Although our findings are consistent with the MDH,
the MDH itself does not explain the volatility dynamics
which may account for other relevant properties such as 
volatility clustering and long-memory of absolute returns. 
In order to understand the complete price dynamics 
more studies needed, especially for volatility dynamics.
In econophysics absolute returns are often used as
a substitute for volatility measures 
which do not match the integrated volatility\cite{VOL1,VOL2,VOL3,VOL4,VOL5}.
Since the realized volatility calculated by using high-frequency data
is an accurate measure of the integrated volatility 
it may serve as a tool for further studies of the volatility dynamics.

\section*{Acknowledgements}
Numerical calculations in this work were carried out at the
Yukawa Institute Computer Facility 
and the facilities of the Institute of Statistical Mathematics.
This work was supported by Grant-in-Aid for Scientific Research (C) (No.22500267).

\end{document}